# Testing the distance-dependence of the van der Waals interaction between an atom and a surface through spectroscopy in a vapour nanocell

A. Laliotis, I. Maurin, P. Todorov, I. Hamdi, G. Dutier, A. Yarovitski,
S. Saltiel, M.-P. Gorza, M. Fichet, M. Ducloy and <u>D. Bloch</u>

Laboratoire de Physique des Lasers, UMR7538 du CNRS et de l'Université Paris13,
99 Avenue Jean-Baptiste Clément, 93430 Villetaneuse, France

## ABSTRACT

This paper presents our current measurements in a vapor nanocell aiming at a test of the distance-dependence of the atom-surface interaction, when simple asymptotic descriptions may turn to be not valid. A state-of-the-art of atom-surface interaction measurements is provided as an introduction, along with the comparison with the theory of the van der Waals (or Casimir-Polder) interaction; it is followed by a presentation of the most salient features of nanocell spectroscopy.

## 1. From van der Waals to Casimir interaction and the state-of-the-art in the measurements

The interaction between neutral bodies can be seen as the universal attraction at the scale of atomic, molecular and solid-state physics. It is indeed at the origin of covalent interaction, and other type of "van der Waals" crystals. If interaction between "particles" was recognized by van der Waals (1873) to be a major cause for gases not to follow the equation of state of an ideal gas, the effective origin of these interactions were identified in the 1930 with the works of London and Lennard-Jones and coined as "van der Waals" (hereafter vW) type (for references, see [1,2]). It is indeed the instantaneous correlation between (quantum) fluctuations of particles that is responsible for their attraction. For an atom in front of a reflecting surface, the concept of electrostatic image implies that an electric image of each of the elementary charges in the atomic system atom is created. Although the atom is neutral, and even without a permanent dipole, the atom is attracted by its image through the non-zero dipole–dipole attraction resulting from its quantum dipole fluctuations. These fluctuations have to be developed over the (virtual) dipole transitions. Naturally, the idea of an "instantaneous" correlation of fluctuation is not acceptable in regards to light propagation, in particular when the distance of separation is non negligible relatively to the wavelength of the considered transition. As early as 1948, the Casimir-Polder work [3], triggered by the "mundane" problem of interaction between colloids, has shown that retardation effect should be considered when the atom-surface distance is non negligible relatively to the wavelength of the virtual transitions responsible for the quantum dipole fluctuations. In particular, the predicted $1/z^3$ potential (z: the atom-surface distance) should be replaced by an interaction decaying even faster with the distance, up to a long range regime in $1/z^4$. In the same year, Casimir discovered as a consequence of Quantum Electrodynamics that two reflecting surfaces should attract each other, with a $1/d^4$ dependent pressure (d : the distance between the two surfaces) [4].

Although a very tiny effect in most cases, the Casimir interaction (for reviews, see [5]) has been the topic of some experiments in the 50's, that were merely able to evidence an attraction compatible with the prediction. In the same time, an alternative theoretical approach was developed by Lifchitz based upon thermodynamics arguments [6]. Some experiments on the vW interaction (see [2] and references therein) between two plates should also be included in these early attempts to measure the Casimir interaction. Only recently the Casimir interaction has become a hot spot of fundamental physics, following the experimental advances of Lamoreaux in 1997 [7] - with a torsion balance- and of Mohideen *et al.* [8] with an atomic force microscope (since 1999), and the claim of very high accuracy ( ~ 1%) and of a succesful comparison between the experiment and the theory. In these experiments, the measurement includes a demonstration of the distance dependence on a 0.6μm-6μm range in the Lamoreaux experiment, in the 100-900 nm range for the Mohideen work, a range that has been now pushed down to 62 nm. These experiments have stimulated refinements in the theoretical description (QED at non zero temperature, corrections for non ideal reflectivity, ... influence of the exact shape of the surface, ... ). It is now recognized that the evaluation of the cosmological constant may depend on the exactness of the QED predictions among which the Casimir interaction is a major test. Also, the measurement of a possible deviation to the Newtonian gravitation ("fifth force") with an additional short distance gravity can be attempted only if the unavoidable Casimir interaction is rightly taken into account [5]. Apart from this fundamental interest, the Casimir interaction being responsible for a pressure comparable to the atmospheric pressure for d~10 nm, is also nowadays considered as a limitation, or a possible tool, for the development of nano-actuators (NEMS).

Measurements of the atom-surface interaction had remained scarce for a long time owing to the extreme weakness of the force. If the vW interaction for an atom-atom potential can be of the order of a few tenths of eV, and can be estimated, but with a limited accuracy, as the asymptotic tail of an interatomic potential for distances exceeding several atomic units, the current distance of observation for atom-surface interaction is considerably larger, leading to extremely tiny effects. Remarkably, the vW interaction is supposed to span over ~10 orders of magnitude in energy on a distance scale that approximately covers the 1nm-1 μm distance. The first observations were conducted in the Columbia group in the late 60's [9]. They demonstrated a deflection of an atomic (and molecular) beam by a cylinder, and an effective impact parameter covering a range ~50-100 nm (a too strong attraction makes the atoms caught by the surface, a too weak one is undetectable). A deflection compatible with a $z^{-3}$ law could be found, that looked to be relatively more satisfying than a $z^{-2}$ law or a $z^{-4}$ law. About 20 years later, a second generation of experiments [10] was conducted by the Yale group based upon the measurement of the transmission rate of a beam of Rydberg atoms through a system of two metal-coated plates. These experiments allowed a precise observation of the distance-dependence in the 500-3000 nm range - distance between plates-, and also evidenced the predicted dependence of the vW interaction with the level of atomic excitation. Indeed, one expects for a state i> a vW interaction governed by a potential: $V_{vW}(|i>) = -\frac{C_3}{z^3}$, with the $C_3$ coefficient depending on the quantum fluctuations of the atom dipole : $C_3 = \frac{1}{16}\langle i|\mathbf{D}^2 + \mathbf{D}_z^2|i\rangle$ [or $C_3 = \frac{1}{12}\langle i|\mathbf{D}^2|i\rangle$ if the atom is isotropic]. A notable point is that for an atom, one has $\langle i|\mathbf{D}^2|i\rangle \neq 0$, in spite of the absence of a permanent dipole (<i|D|i>=0) : the dipole fluctuations, of quantum origin, corresponds to a summing over all virtual dipole transitions $\langle i|\mathbf{D}^2|i\rangle = \sum_i \langle i|\mathbf{D}|j\rangle^2$, which are much stronger in the far IR, as found for the couplings relevant for a Rydberg level, than for a ground state or a low excited level. In these experiments on Rydberg levels, that combine a mechanical detection with a spectroscopic excitation, energy shifts up to 500 MHz were obtained. A derived set-up was used to obtain the first evidence of the retarded Casimir-Polder regime [11], that can be seen only for ground state atoms, because excited atoms tend to undergo a real - and not virtual- emission process: indeed, excited states radiate according to the far field regime of an oscillating dipole, and largely exceeds the retarded vW attraction.

Contemporarily to the Yale's group experiments, our Paris13 group entered into the field of measurement of vW atom-surface interaction [12,13], through a spectroscopic technique (for a review, see [1]) of selective reflection in a vapor that typically probes the vapor on a depth ~λ/2π (~ 100 nm) from the surface. Two peculiar features should be mentioned for these types of experiments: (i) spectroscopy is intrinsically sensitive to the energy of a *transition* between atomic levels, so that what is measured is the difference between the vW interaction for an excited level, and the vW level of another level (ground state for a one-photon experiment, another excited level on more complex set-up); (ii) as a major difference with mechanical detection, spectroscopy allows to study the vW interaction undergone by short-lived atoms, instead of being intrinsically limited to ground states, or long-lived Rydberg levels. Naturally, reflection spectroscopy at an interface requires transparent surfaces, so that *dielectric* materials are investigated. Among the variety of results [1] that we have demonstrated on the basis of this selective reflection technique, one should mention as far as the atom-surface interaction is concerned: (i) the observation of weak and strong vW regimes [13] that bring respectively either a perturbation (shift and distortion) to the lineshape expected in the absence of the vW interaction, or a severe change in the lineshapes, and (ii) the influence of the nature of the dielectric surface, with the possibility to resonantly enhance the vW attraction, or to turn it into a vW repulsion when the surface mode structure coincides with a virtual atomic emission [14]. As will be reported in the next sections, we have essentially upgraded our selective reflection spectroscopic technique, providing a fixed spatial resolution, to a technique of spectroscopy in a vapor nanocell, allowing us to study the spatial dependence of the vW interaction in a sub-100 nm range.

To complete this description of the state-of-the art, it is needed to mention the recent blossom of experimental techniques for the measurement of the vW attraction, that has notably been triggered by the importance of the problem of surface interaction for the storage and manipulation of cold atoms, and by development in atom interferometry as well. Cold atom techniques have been used to measure the vW interaction through a combination with an applied repulsive potential [15]. Very recently, the repulsion from a strong magnetic field allowed a test of the distance dependence in the 20-100 nm range (for Cs ground state [16]), and showed a $C_3$ value remaining constant within 15-20%. Interestingly, this verification of a $C_3$ value remaining constant with the atom-surface distance seems to disagree with refined theoretical predictions [17], because at the shorter distances, the transitions from the atomic core, often negligible due to the severe retardation effects owing to their short wavelengths, cannot be ignored. In the other measurements based upon a cold atom technology, it is rather the long distance range that is conveniently probed, as can be inferred from the comparison between the thermal energy $k_B T$ and the vW potential $V_{vW} = -C_3 z^{-3}$. Indeed, for a ground state atom, the vW interaction exceeds the thermal energy only for $z < 1$nm at room temperature, but for $z \sim 100$ nm one has already an interaction potential $\sim 300$ μK, corresponding to a temperature much hotter than the one commonly attained with cold atoms. This long-distance probing is a favorable situation for the observation of the Casimir-Polder regime, probably observed in [15], and that influences the quantum reflection of very cold atoms [18]. A record long distance -6μm- has now been attained in a BEC-type experiment [19]. This paves the way to observation of temperature effects, and enables to rule out some coupling strength for an hypothetical non-Newtonian gravity whose range would span over the micronic range. In addition to these long-distance measurements enabled with cold atoms, experiments connected to atom interferometery [20] enable the exploration of the interaction at a very short distances (~5 nm), in a manner that globally integrates the spatial variation of the vW potential, not allowing to resolve details concerning its shape or spatial dependence.

## 2. Spectroscopy in micro- and nano-cells of resonant vapor : from sub-Doppler spectroscopy to atom-surface interaction

Selective reflection spectroscopy under near normal incidence has been known from early times [21] to lead to sub-Doppler resonances, recognized to originate in the transient interaction regime undergone by those atoms leaving the surface. Indeed, in selective reflection, only the atoms of a slice of vapor of a thickness $\lambda/2\pi$ emit backward coherently, so that under normal incidence, only slow atoms (insensitive to the Doppler shift) reach an efficient steady-state of interaction in the ascribed length. This sub-Doppler nature of selective reflection spectroscopy, that helps to make this technique sensitive to the atom-surface interaction [1] - because the response of atoms spending a long time in a region of strong atom-surface interaction is enhanced- has inspired us to look for an equivalent signature in the transmission regime [22,23], expected to be independent of the cell length – as opposed to regular volume absorption. Our earliest experimental attempts were conducted on 10-1000 μm thin vapor cells [24, 25], in order to maximize the ratio between the region sensitive to atoms departing from the surface, and the isotropic "bulk" of the vapor. The pressure of the (dilute) vapor was kept low enough so that most atomic trajectories are wall-to-wall, with the light beam diameter largely exceeding the cell thickness. This offers the key condition of an atomic free path that becomes anisotropic. With such microcells, we observed indeed a sub-Doppler signal (whose contrast is enhanced with a frequency derivation technique, as provided by demodulating an applied FM), yielding an original sub-Doppler method with a single beam, that originated in a velocity-selective non linear effect (enhanced optical pumping for slow atoms). Such a sub-Doppler spectroscopic technique had been independently predicted by Izmailov [26]. Soon after, in an effort to keep incident intensities well below the (quite low) intensities required for optical pumping, we obtained in a linear regime a different sub-Doppler signature [27], purely relying on a two-level model (with no need of considering an "open" two-level model, *i.e.* a model where at least a third level influences the atomic lifetime). Several differences can be pointed out with the sub-Doppler regime of optical pumping : in particular, the sub-Doppler signature, even in FM technique, is reduced if the cell length exceeds the mean distance needed for a thermal atom to reach the steady state. Also, because 2-level absorption is a coherent process, the sub-Doppler spectral properties, initially calculated in [23] for a situation limited to a thickness between $\lambda/2$ and $\lambda$, can be shown to depend interferometrically on the cell length, in a way consistent with the predictions by Romer and Dicke [28], established 50 years ago for microwave experiments : for a $\lambda/2$ thick cell, the transient response of all velocity groups interfere constructively at line center, yielding a narrow signal, while for a $\lambda$-thick cell, the successive absorption and gain corresponding to a full Rabi oscillation would lead to a null signal in the absence of absorption relaxation. This is why successive revivals – with a $\lambda$ periodicity- of these spectroscopic features is predicted [25,27]. If some interesting features for sub-Doppler spectroscopy could be revealed with experiments in a 10 μm-thick cell, such a cell was nevertheless not convenient to observe the vW atom-surface interaction.

As will be established in the next section, the production of sub-micrometric thin vapor cells by the D. Sarkisyan group [29] has been the key to apply realistically spectroscopy in a thin cell to the monitoring of the atom-surface interaction. A remarkable point with these cells, that are described in more detail in the presentation given by D. Sarkisyan [30], is that after evacuation (before proceeding to the filling with a metallic vapor), and in spite of their construction with two thick and well-polished windows separated by a spacer of a constant thickness, the cell undergoes local deformations of the windows, leading to a variable local thickness of the thin cell, usually spanning from a few 10's of nm to one or a few microns [31]. This provides a key ingredient, when using a focused beam, to study the thickness dependence of the spectrum and ultimately to monitor the distance dependence of the vW interaction. Practically, it is indeed sufficient to simply translate the cell, that is truly a "nanocell" in some regions. Before being able to perform an evaluation of the atom-surface potential, it has been needed to demonstrate a full understanding of the spectroscopy in these sub-micrometric cells. A notable point is that, because, the two windows of a nanocell are intrinsically extremely parallel – in spite of the minute deformation after evacuation- a nanocell is a genuine Fabry-Perot cavity [32], although of a low finesse : on the one hand, this provides a very suitable tool to measure the local cell thickness with an interferometric accuracy [31]; on the other hand; all nanocell spectra are unavoidably the result of a mixing [32] between absorption-like transmission, and reflection (which would be a pure dispersive signal only for an infinite cell length (see appendix of ref. 13). Our initial studies have enabled the observation of the Dicke revival as predicted in [25,27] with the corresponding lineshapes both in transmission and reflection detection. The easiest interpretations are naturally obtained in non saturating conditions, that are more convenient to reach than in longer cells (even micro-cells): indeed the interaction time is very limited, and the smaller interaction region is usually compensated by a stronger atomic density, implying an increase in the saturation intensity because of collision broadening. Note that a fluorescence detection can be used in nanocell spectroscopy, as it eventually provides narrower linewidths than transmission/reflection spectra [29,33]. However, fluorescence is an *incoherent* second-order process: apart from being insensitive to the periodical *coherent* Dicke narrowing [33], this makes fluorescence spectra intrinsically much more delicate to interpret when the evaluation of the atom-surface interaction is at stake.

## 3. The monitoring of atom-surface interaction in nanocells

The general theory of spectroscopy in a thin cell or a nanocell [32] resembles by many aspects the theory of selective reflection spectroscopy in the presence of a potential, as developed in [34]. There are additional spectroscopic effects due to the finite length of the cell (*i.e.* integrations are performed within finite limits, and not within infinite boundaries [22]), and eventually the Fabry-Perot mixing of transmission and reflection. Moreover, a dual detection of reflection and transmission is possible. The basic ingredients of the theory include :

 (i) evaluation of the atomic response at a given point, as resulting from the integration of the transient response from the departing wall to the actual position. This response, that would be elementary to calculate - and time integrate- in the absence of a vW potential, depends on the interaction potential (for a 2 wall-interaction). An interesting variety of situations has to be considered if the dispersive nature of the surface response is included, that implies a lot of complexity to describe the shape of the interacting potential [35, 36]. In a pure electrostatic approach, that neglects the dispersive properties of the surface, the potential should include the equivalent of multiple electrostatic images, and the interacting potential is then described with a Lerch function. Although not rigorous, an elementary model simply adding the elementary vW contribution of each wall (*i.e.* $V(z) = -C_3[(1/z)^3 + [(1/d-z)^3]$, with d the cell thickness) can be most often sufficient, as the difference with the exact Lerch potential does not exceed 5 % when dispersive properties are neglected.

  (ii) spatial integration all over the cell of the instantaneous response of an atom at a given position; for the reflective contribution, or in some terms of the transmission response when a Fabry-Perot-like response has to be considered, this integration has to includes an *exp*$(2ikz)$ phase factor (with k the wave vector).

  (iii) a velocity integration, over the (usually) thermal distribution of atoms. Instead of a family of curves depending on a unique parameter that describes in dimensionless units all the features of vW, as was obtained for selective reflection spectroscopy (at least in the approximation of the infinite Doppler width), one predicts for thin cell spectroscopy independent sets of spectral lineshapes when varying the cell thickness, or the vW strength, or also the velocity distribution (in a regime of small thickness, the vW interaction is not negligible relatively to the Doppler broadening, and the "infinite Doppler width" cannot hold).

 Figure 1 illustrates some predicted lineshapes, when the strength of the van der Waals is varied for a given thickness, and when the cell thickness varies for a given vW coefficient. The potential is an approximate Lerch function, but similar results would actually be obtained for the elementary summing of two single-wall potentials. Experimentally, we have already mentioned in [36] that by using the above described modeling of the transmission lineshapes in a nano-cell, we observe on the Cs $D_1$ resonance line ($\lambda$=894 nm) some deviations relatively to the lineshape predicted in the absence of an atom-surface interaction for a 225 nm ($\lambda$/4)-thick cell. Also, in even earlier experiments [37], we had shown through a stepwise two-photon excitation that allows to reach the high-lying 6D level of Cs, that considerable shifts and broadenings could be observed in extremely thin regions. We had even attempted to verify a $z^{-3}$ dependence through the position of the transmission peak. However, these preliminary experiments were not conducted in a systematic way: the pressure effects were not controlled, and there was at that time no lineshape analysis. Indeed, the major goal was an attempt to check for resonant 2-wall atom-surface coupling, as our observation of a vW repulsion had been performed on Cs($6D_{3/2}$) in front of a sapphire surface [14].

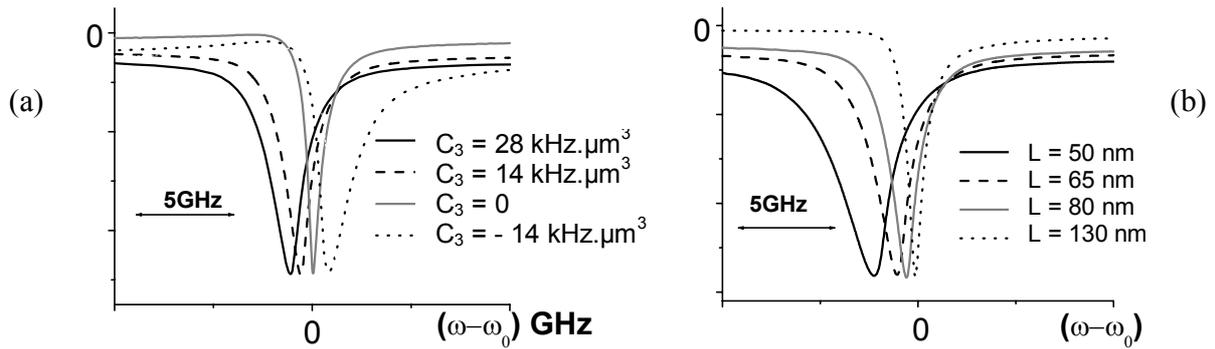

Figure 1 : Theoretical transmission lineshapes for a Doppler width ku = 250 MHz :
(a) for a 80 nm thickness and various $C_3$ values, (b) for $C_3$ = 14 kHz.µm$^3$ and indicated thicknesses.

Here our report is on a systematic series of experiments [38] performed on the Cs($6P_{3/2}$)-Cs($6D_{5/2}$) transition ($\lambda$= 917 nm), as reached after a pumping step ($\lambda_{pump}$= 852 nm) to the resonant Cs($6P_{3/2}$) level. Such an experimental choice has two important conveniences : on the one hand, the detection sensitivity approaches the quantum noise sensitivity in a modulated pump-probe scheme, as the detection is not performed at a null frequency (when all low-frequency noises are present), but at the frequency of the modulation applied to the pump (50 kHz in our case) : the sensitivity issue is obviously an important one when the measurements of interest should deal with the minimal atomic density; on the other hand, the vW interaction is predicted to be larger by about an order of magnitude for Cs(6D) than for Cs(6P), so that larger interactions should be observed if there is a minimal nanocell thickness that can be technically reached. The principle of the experiment, schematized in figure 2, is very simple. One monitors, through lock-in detectors, the changes in the transmission (T) and reflection (R) signals in the nanocell when the pump beam is on. The incident pump and probes are focused (diameter~100 µm). The pump and probe beam are nearly counterpropagating (to avoid unwanted pump scattering light) and sent at normal incidence on the nanocell. The local nanocell thickness is measured through the reflection coefficient of the pump beam (once detuned off-resonance) and of an auxiliary He-Ne beam : in the range of very small thickness ( <150 nm) that we consider, a dual wavelength measurement is sufficient to provide unambiguously a reliable thickness measurement. Moreover, being interested in the smaller thickness, we explore a region of the cell that is intrinsically flat, and the accuracy in the thickness measurement, currently ~ 5 nm, can occasionnally reach a 1-2 nm level. An auxiliary reference set-up of volume saturation spectroscopy is implemented on the 917nm line (with prior 852 nm pumping), while the 852 nm frequency itself is kept fixed on one hyperfine component of a Doppler-free saturated absorption line.

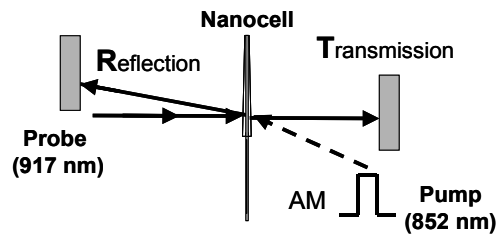

Figure 2 : principal scheme of the experimental set-up

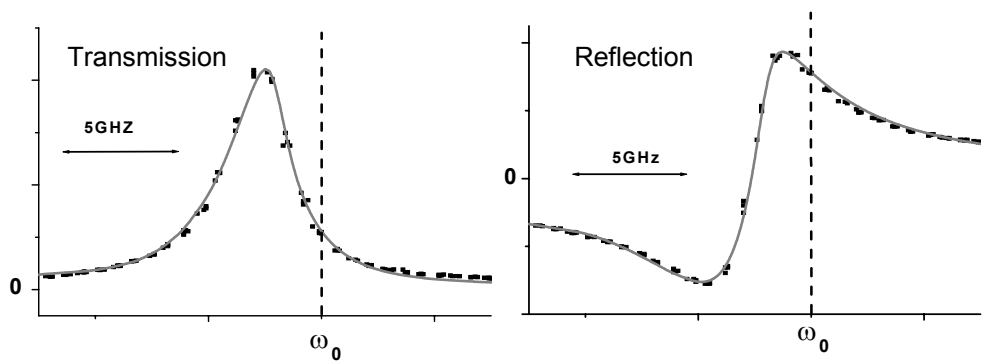

Figure 3 : Transmission and reflection spectra recorded in a 50 nm nanocell, with a discreetly tunable laser 917 nm laser. The full lines are a theoretical fitting. The vertical dashed line is a frequency reference as determined by an auxiliary saturated absorption (volume) reference

A typical experimental result is shown in figure 3 for a thickness 50 nm, with a simultaneous recording of transmission (T) and reflection (R) spectra. The spectra are quite broad (the Doppler width is only ~250 MHz) and strongly red-shifted relatively to the volume reference. The influence of the pressure on these features reveals to be quite weak (*e.g.* spectra recorded with a higher Cs pressure remain unchanged). Such spectra can be fitted with one set (T and R) of the calculated lineshapes for d=50 nm. In the fitting adjustment, only adjustable amplitudes have been used (one parameter for T, one for R) . In the velocity integration, the resonant $Cs(6P_{3/2})$ level has been assumed to be pumped in a thermal manner. Even if such an assumption may be problematic, the influence of this velocity distribution on the global spectrum remains quite weak - in particular, the choice of a smaller Doppler width, that would correspond to a more efficient pumping of the slow atoms, provides very similar results-. A notable result is hence that T and R lineshapes can be fitted simultaneously with the same $C_3$ parameter. In this regime in which the vW interaction dominates over the pressure or Doppler broadening, it is easily seen (see *e.g.* fig . 1) that any attempt to fit a given lineshape with notably different $C_3$ values will lead to a differing frequency shift at the peak, and to different lineshapes. It is only for marginal changes of $C_3$ around an acceptable value that a least square fitting is a help to determine the $C_3$ value with an increased accuracy.

To study systematically the distance dependence of the $C_3$ value, shown above to be measurable with our technique, we had studied various Cs densities (pressure) conditions for each considered cell thickness. As shown in figure 4, we obtain constant $C_3$ values independent of the pressure, except for the larger thickness, when the vW interaction can turn out to be smaller than the pressure broadening – for high Cs densities, or temperature, as an increase of 20°C is roughly a doubling in Cs density- Also, for the larger thickness, the unresolved hyperfine structure of the pumped state Cs ($6P_{3/2}$) - 350 or 450 MHz depending on the ground state hyperfine component - makes more difficult the frequency referencing to the volume resonance. This constant value of the estimated $C_3$ vW interaction constant is illustrated in the spectra of figure 5, when the conditions for Cs vapor are identical, and when a single set of parameter for $C_3$ and for the transition width is considered. Only the amplitudes are adjustable by an arbitrary factor, because the density of active Cs – in the resonant $6P_{3/2}$ level - is expected to change with the thickness, owing to the limited time available for the excitation. Also, for a given thickness, slight disagreements between the relative amplitudes for R and T can be simply due to the (local) scattering properties of the surfaces, which makes the non resonant reflection and transmission factor apparently disagree with the Fresnel formulae.

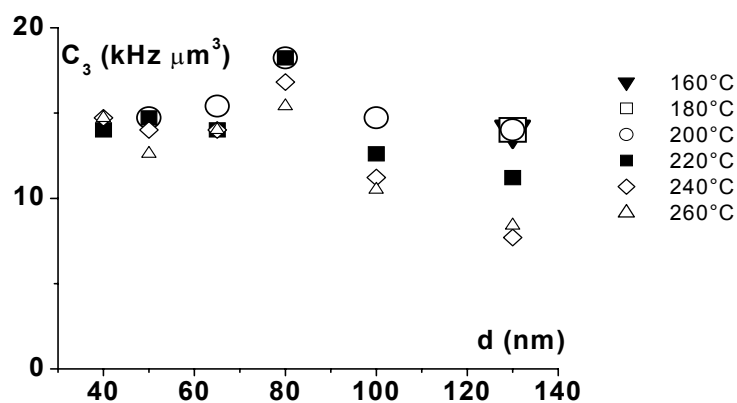

Figure 4 : optimal fittings $C_3$ values for various thicknesses, and in differing pressure conditions

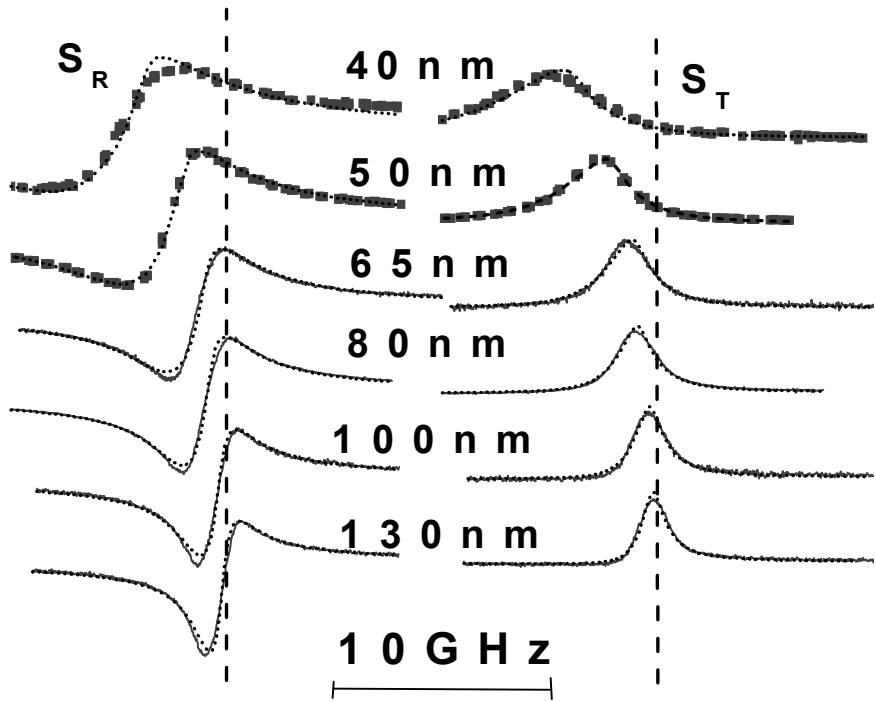

Figure 5 : Reflection and Transmission 917nm spectra recorded in a Cs nanocell heated-up at 200°C for different local thickness. The dotted lines are a fitting obtained with a single $C_3$ value 14 kHz.$\mu m^3$, and a single width for the optical transition 300 MHz. The vertical dashed line is a reference frequency for the volume transition

Before comparing the experimental findings and the predictions, it is worth discussing the reproducibility of the results presented here. The surface that we probe is typically a spot ~(100μm)² size, as a result of the laser focusing. Although this is a small size relatively to most comparable studies, this remains very large in comparison to the atom-surface distance. The defects in planarity, roughness, or parallelism of the surfaces, naturally affect the measurement of the distance-dependence that we are interested in. Moreover, it is known that various stray fields can severely affect the measurement of atom-surface interaction, notably for a dielectric surface. Indeed, local domains generate residual fields. This is why we have repeated each measurement performed on a spot of a given "nominal" (average) thickness (as determined by our 2-wavelength interferometric measurement) by measurements on other spots of a similar nominal thickness, either through a systematic density analysis, or at least in the most significant conditions (*i.e.* low density). We have found that for all experiments for d ≥ 65 nm or above, the spectra are identical within our accuracy. For d = 50 nm, some variations are observed from spot to spot, although some spots are able to provide identical spectra. These spectra only are retained, because any broadening or shift induced by a residual field would be strongly specific of the considered spot, not being susceptible to produce the same spectrum as another spot. For d = 40 nm, the vW shift is so considerable that minor changes (*e.g.* in the distribution of the local thickness over the spot size) can induce visible changes on the lineshape. However, the analysis of the lineshapes reveal that these minute changes have a negligible influence on the parameters that can be estimated in the fitting procedure. This reproducibility may appear unexpectedly good relatively to what can be inferred from other experimental results. Among the favorable points in our experiments, is the fact that the vW energy is pretty strong, owing to the investigated range of small distances, and to the relatively low excitation of the Cs (6D) (notably in comparison with experiments on Rydberg levels that are very sensitive to residual Stark shifts)

## 4. Discussion and prospects

Our experiments evidence the possibility to fit the lineshapes for the Cs($6P_{3/2}$)-Cs($6D_{5/2}$) transition with lineshapes governed by a $C_3$ value ~ 14 (±3) kHz/µm$^3$ for a cell thickness comprised between 40 nm and 130 nm, and hence for a maximal atom-surface distance ranging from 20 nm to 65 nm. This corresponds to a surface-induced shift obeying a $1/z^3$ law for an energy range spanning from ~200 MHz to 6 GHz, *i.e.* a factor ~30. Higher energies have been demonstrated to be observable [36,37], although not fully analyzed, and hence perhaps not to be attributed entirely to the surface interaction. Nevertheless, these experiments represent a bridge between the asymptotic limit of long distance Casimir-polder measurements, and the tail of a chemical potential. It shows also that in spite of a strong interacting potential, the atomic structure is still recognizable in spite of important shifts and broadening. Nevertheless, it is clear that in the explored range, with the observation of record energy shifts exceeding several GHz, mechanical effect should have considerable effects for the slowest atoms.

In addition to the experimental findings, the comparison with the theoretical prediction, detailed in [38], is remarkably good. The theoretical predictions is indeed evaluated to ~15 kHz./µm$^3$ [– *i.e.* 17 kHz./µm$^3$ for Cs(6D), and 2 kHz./µm$^3$ for Cs(6P)], once taken into account the dielectric coefficient, and a summing over the virtual transitions, assumed to be isotropically pumped. There are anyhow various pitfalls susceptible to affect the theoretical predictions: the modeling assumes the population to be thermal and spatially homogeneous, while it is here provided through a pumping step in a confined environment. The modeling also assumes an isotropic preparation of substates, allowing all orientations of the fluctuating atomic dipole to contribute [35]. The interaction between the highly-excited Cs atom, and the YAG windows, may include some resonant contributions, hard to determine in view of various uncertainties concerning the detailed spectrum of YAG surface resonances [39].

Preliminary experiments on the Cs $D_1$ resonance line ( 894 nm) are now in progress. Their interpretation, and the possible comparison with theory, should be easier than for an interaction mostly affecting a highly-excited state. However, the additional difficulties are that shorter distances are needed to reach a comparable strength of the vW interaction. Also, the sensitivity is affected by the fact there is no possibility to modulate an external (pump) beam, so that the detection has to be operated with all the amplitude noise of the source in the low-frequency range. Preliminary measurements enable us to record reproducible spectra down to a 30 nm thickness - *i.e.* all atoms are at less than 15 nm from the wall- but the observed vW shift seems larger (by a factor up to 2) than expected.

Our measurements clearly show that reaching an effective atom-surface distance below 10 nm is a realistic objective. This contrasts with the measurements based upon the reflection of slow atoms [15,16] when the minimal distance of approach of the observed atomic trajectories is limited by the considerable force needed to compensate for the vW attraction. In the range of small distances that we explore here, we have evidenced an interaction energy largely exceeding the ones obtained for all previous investigations (up to ~5 GHz, or 0.25K). The corresponding acceleration is considerable (~ 8.10$^7$ g for a Cs(6D) atom 20 nm away from one of the wall), notably with respect to the one obtained in laser cooling techniques. Extrapolating to even smaller distance, one may even wonder if some kind of "atmospheric" gradient of density could not be found for an atomic gas in the extreme vicinity with a surface. The specificity of a spectroscopic measurement implies that we are essentially sensitive to a difference between the interaction undergone by the different states: on the one hand, this may make our measurements insensitive to an hypothetical non Newtonian gravity, on the other hand, this allows a simpler test of the validity of the van der Waals interaction at small distances, nearly-free of the influence of core transitions, as they are expected to be nearly independent of the excitation degree of the peripheral electron [1]. At last, our measurements on excited states could be pushed up to the situation of resonantly coupled excited states, with atoms confined between two walls by a repulsive potential [36].

## Acknowledgments

Work supported by the European Union through the research training network FASTnet (contract HPRN-CT-2002-00304), and by the French-Bulgarian programme of co-operation (RILA 09813UK)

# REFERENCES

1. D. Bloch and M. Ducloy "*Atom-wall interaction*", Adv. At. Mol. Opt. Phys., **50** pp. 91-154 (B. Bederson and H. Walther eds., Elsevier-Academic Press, 2005).
2. J.N. Israelachvili and D.Tabor, Proc. Roy. Soc. A, **331**, 19-38 (1972) and references therein
3. H. B. G. Casimir and D. Polder. Phys. Rev., **73**, 360 (1948)
4. H. B. G. Casimir, Proc. Kon. Ned. Akad. Wetenshap **60**, 793 (1948)
5. M. Bordag, U. Mohideen and V. Mostepanenko, Phys. Rep. **353**, 1 (2001); A. Lambrecht and S. Reynaud, in *Poincaré seminar 2002*, *Vacuum energy*, (B.V. Rivasseau ed., Birkhauser,) p.109 (2003); K.A. Milton, J. Phys. A, **37**, R209 (2004).
6. E. M. Lifshitz. Zh. Eksp. Teor. Fiz., **29**,94, 1956. [Soviet Phys. JETP **2**, 3,(1956)]
7. S. K. Lamoreaux, Phys. Rev. Lett. 78, 5 (1997); e : **81**, 5475 (1998)
8. U. Mohideen and A. Roy, Phys. Rev. Lett. **81**, 4549 (1998); see also F. Chen, G. L. Klimchitskaya, U. Mohideen and V. M. Mostepanenko, Phys. Rev. A **69**, 022117 (2004)
9. D. Raskin and P. Kusch, Phys. Rev. **179**, 712 (1969; A. Shih and V.A. Parsegian *Phys. Rev. A* **12**, 835 (1975).
10. V. Sandoghdar, C. I. Sukenik, E. A. Hinds, and S. Haroche V.*,* Phys. Rev . Lett. **68** 3432 (1992); V. Sandoghdar, C. I. Sukenik, S. Haroche, and E. A. Hinds, Phys. Rev. A **53**, 1919 (1996).
11. C. I. Sukenik, M. G. Boshier, D. Cho, V. Sandoghdar, and E. A. Hinds, Phys. Rev. Lett. **70**, 560 (1993)
12. M. Oriá, M. Chevrollier, D. Bloch, M. Fichet and M. Ducloy, Europhys. Lett. **14**, 527 (1991)
13. M. Chevrollier M. Fichet, M. Oriá, G. Rahmat, D. Bloch and M. Ducloy, J. Phys. II *(France) 2*, 631(1992)
14. H. Failache, S. Saltiel, M. Fichet, D. Bloch and M. Ducloy*,* Phys. Rev. Lett. **83**, 5467 (1999); Eur. Phys. J. D, **23**, 237 (2003).
15. A. Landragin, J.-Y. Courtois, G. Labeyrie, N. Vansteenkiste, C.I. Westbrook and A. Aspect, Phys. Rev. Lett. **77**, 1464 (1996).
16. A. K. Mohapatra and C. S. Unnikrishnan, Europhys. Lett*.*, **73,** 839 (2006)
17. A. O. Caride, G. L. Klimchitskaya, V. M. Mostepanenko and S. I. Zanette, Phys. Rev. A **71**, 042901 (2005).
18. F. Shimizu, Phys. Rev. Lett. **86**, 987 (2001)
19. D. M. Harber, J. M. Obrecht, J. M. McGuirk and E. A. Cornell, Phys. Rev. A **72**, 033610 (2005)
20. R.E. Grisenti, W. Schöllkopf, J.P. Toennies, G.C. Hegerfeldt and T. Köhler, Phys. Rev. Lett. **83**, 1755 (1999) M. Boustimi, B. Viaris de Lesegno, J. Baudn, J. Robert and M. Ducloy, Phys. Rev. Lett., **86**, 2766 (2001).
21. M.F.H. Schuurmans, J. Phys.(Paris) **37**, 469 (1976), see also J.-L. Cojan, Ann. Phys. (Paris*) 9*, 385 (1954).
22. T. A. Vartanyan and D. L. Lin, Phys. Rev. A **51**, 1959–1964 (1995)
23. B. Zambon and G. Nienhuis, Opt. Commun. **143**, 308 (1997).
24. S. Briaudeau, D. Bloch and M. Ducloy M., Europhys. Lett*.* **35,** 337 (1996); Phys. Rev. A **59**, 3723 (1999).
25. S. Briaudeau, PhD thesis, Université Paris 13 (1998) (unpublished).
26. A. Ch. Izmailov, Laser Phys*.*, **2**, 762 (1992); **3**, 507(1993); Opt. Spectrosc., **74**, 25 (1993); **75,** 395 (1994).
27. S. Briaudeau, S. Saltiel, G. Nienhuis, D. Bloch, and M. Ducloy, Phys. Rev. A **57**, R3169 (1998).
28. R. H. Romer and R. H. Dicke, Phys. Rev. **99**, 532 (1955).
29. D. Sarkisyan, D. Bloch, A. Papoyan and M. Ducloy, Opt. Commun. **200**, 201 (2001).
30. D. Sarkisyan, see contribution in the same SPIE Proceedings volume
31. G. Dutier, A. Yarovitski, S. Saltiel, A. Papoyan, D. Sarkisyan, D. Bloch and M. Ducloy, Europhys. Lett. **63**, 35 (2003)
32. G. Dutier, S. Saltiel, D. Bloch and M. Ducloy. J. Opt. Soc. Am. B, **20**, 793 (2003)
33. D. Sarkisyan, T. Varzhapetyan, A. Sarkisyan, Yu. Malakyan, A. Papoyan, A. Lezama, D. Bloch and M. Ducloy, Phys. Rev. A **69**, 065802 (2004)
34. M. Ducloy and M. Fichet, J. Phys. II **1**, 1429 (1991).
35. M.-P. Gorza *et al*, in preparation
36 I. Hamdi , P. Todorov, A. Yarovitski, G. Dutier, I. Maurin, S. Saltiel, Y. Li, A. Lezama,T. Varzhapetyan, D. Sarkisyan, M.-P. Gorza, M. Fichet, D. Bloch and M. Ducloy, Laser Phys, **15**, 987 (2005).
37. G. Dutier, A. Yarovitski, S. Saltiel, D. Sarkisyan, A. Papoyan, T. Varzhapetyan, D. Bloch, M. Ducloy, in "*Laser Spectroscopy, Proceedings of the XVI International Conference*", (P. Hannaford *et al*., eds., World Scientific, Singapore, 2004) pp. 277-284
38. M. Fichet, G. Dutier, A.Yarovitsky, P. Todorov, I. Hamdi, I. Maurin, S. Saltiel, D. Sarkisyan, M.-P. Gorza, D. Bloch and M. Ducloy., submitted to Europhys. Lett (http://hal.ccsd.cnrs.fr/ccsd-00069106)
39. S. Saltiel, D. Bloch and M. Ducloy, Opt. Commun. **265**, 220 (2006)